\documentclass[12pt]{iopart}

\usepackage[final]{epsfig}

\begin{document}

\title[Elastic Energy Loss of Hard Partons]{Monte Carlo Simulation for Elastic Energy Loss of Hard Partons in a Hydrodynamical Background}

\author{\underline{J.~Auvinen}, K.~J.~Eskola, H.~Holopainen and T.~Renk}

\address{Department of Physics, P.O. Box 35, FI-40014 University of Jyv\"askyl\"a, Finland}
\address{Helsinki Institute of Physics, P.O. Box 64, FI-00014 University of Helsinki, Finland}
\ead{jussi.a.m.auvinen@jyu.fi}

\begin{abstract}
We have developed a Monte Carlo simulation describing the $2 \rightarrow 2$ scatterings of perturbatively produced, non-eikonally propagating high-energy partons with the quarks and gluons of the expanding QCD medium created in ultrarelativistic heavy ion collisions. The partonic scattering rates are computed in leading-order perturbative QCD (pQCD), while three different hydrodynamical scenarios are used to model the strongly interacting medium. We compare our results and tune the model with the neutral pion suppression observed in $\sqrt{s_{NN}}=200$ GeV Au+Au collisions at the BNL-RHIC. We find the incoherent nature of elastic energy loss incompatible with the measured angular dependence of the suppression. The effects of the initial state density fluctuations of the bulk medium are found to be small. Also the extrapolation from RHIC to the LHC is discussed.
\end{abstract}

\section{Introduction} 
Our Monte Carlo (MC) simulation \cite{Auvinen:2009qm,Auvinen:2010yt,Renk:2011qi} aims to study the pQCD-based collisional energy loss of hard partons traversing a strongly interacting medium. It is similar to the MC models JEWEL (Jet Evolution With Energy Loss) \cite{JEWEL} and MARTINI (Modular Algorithm for Relativistic Treatment of heavy IoN Interactions) \cite{MARTINI}, but while JEWEL and MARTINI include both elastic and radiative energy-loss components, we concentrate purely on the energy loss induced by the elastic $2 \rightarrow 2$ scattering processes. In this study, we aim to answer the following questions: i) Within reasonable parameter values, how large could the elastic energy loss contribution be? ii) Does the pathlength dependence of elastic (or, in general, incoherent) energy loss match the experimental data? iii) What are the effects of initial state density fluctuations to energy loss?

\section{The model} 
We model the elastic energy loss of a hard parton by incoherent partonic $2\rightarrow$ 2 processes in pQCD with scattering partners from the medium, upgrading our hydro as we move to the more complex observables. For central heavy ion collisions, a (1+1)-dimensional hydro \cite{Eskola:2005ue} with initial conditions from the EKRT model \cite{EKRT} is used to describe the medium. For non-central collisions, we utilize a (2+1)-dimensional hydro \cite{Holopainen:2010gz} with a smooth sWN profile \cite{Kolb:2001qz} obtained from the optical Glauber model. Finally, to study the effects of the initial state density fluctuations, we have an event-by-event hydro \cite{Holopainen:2010gz} with an eBC profile \cite{Kolb:2001qz} from the Monte Carlo Glauber model. Additionally, we use (2+1)-dimensional hydro with a smooth eBC profile \cite{Renk:2011gj} for LHC simulations.

The basis of our approach is the scattering rate $\Gamma_i (p_1,u(x),T(x))$ for a high-energy parton of a type $i$ with 4-momentum $p_1$, accounting for all possible partonic processes $ij\rightarrow kl$. The hydrodynamical model provides the local flow 4-velocity $u(x)$ and temperature $T(x)$ of the medium. In the local rest-frame of the fluid, we can express the scattering rate for a process $ij\rightarrow kl$ as follows \cite{Auvinen:2009qm}:

\begin{equation*}
\label{scattrate}
\Gamma_{ij\rightarrow kl} = \frac{1}{1+\delta_{kl}} \frac{1}{256\pi^3E_1^2}\int_{\frac{m^2}{2E_1}}^{\infty}dE_2f_j(E_2,T) \int_{2m^2}^{4E_1E_2} \frac{ds}{s} \int_{-s+m^2}^{-m^2} dt \, |M|_{ij\rightarrow kl}^2.
\end{equation*}
Here $E_1$ is the energy of the high-energy parton $i$ in this frame and $E_2$ is the energy of the thermal particle $j$ with a distribution function $f_j(E_2,T)$, which is assumed to be the Bose-Einstein distribution for gluons and the Fermi-Dirac distribution for quarks. The scattering amplitude $|M|_{ij\rightarrow kl}^2$ depends on the standard Mandelstam variables $s,t,u$. A thermal-mass-like overall cut-off scale $m=s_m g_sT$ is introduced in order to regularize the $t,u$ -singularities appearing in the scattering amplitude. The mass factor $s_m$ and the strong coupling constant $\alpha_s = \frac{g_s^2}{4 \pi}$ are the free parameters of our model. We keep $\alpha_s$ fixed with momentum scale.

The hard parton is propagated through the plasma in small time steps $\Delta t$. At each step, the probability for a collision is given by the Poisson distribution $1-e^{-\Gamma_i \Delta t}$. After scattering, the parton with higher energy is selected as the hard parton. As we assume no significant interactions between the high-energy parton and the fully hadronic medium, no collisions are allowed in regions with temperature below the decoupling temperature $T_{\rm dec}$. The medium-modified distribution of high-energy partons obtained in the end can be convoluted with a fragmentation function to calculate the standard nuclear modification factor $R_{AA}(P_T,y,\phi) = \frac{dN_{AA}/dP_Tdy d\phi}{\langle N_{\rm BC}/\sigma_{NN} \rangle\,d\sigma^{pp}/dP_Tdy d\phi}$ for a given centrality.

\section{Results} 
We focus first on the high-$P_T$ neutral pions produced in $\sqrt{s_{NN}}=200$ GeV Au+Au collisions. The simulations for 0-10\% central collisions with the (1+1)-dimensional hydro background show that roughly the right amount of nuclear modification in the most central collisions can be achieved by setting $\alpha_s = 0.5$ and $s_m=1.0$ (see also the left panel of Figure 1). We keep these values as we proceed to non-central collisions with the (2+1)-dimensional hydro to examine the pathlength dependence.

\begin{figure*}[htb]
\centering
\includegraphics[width=16cm]{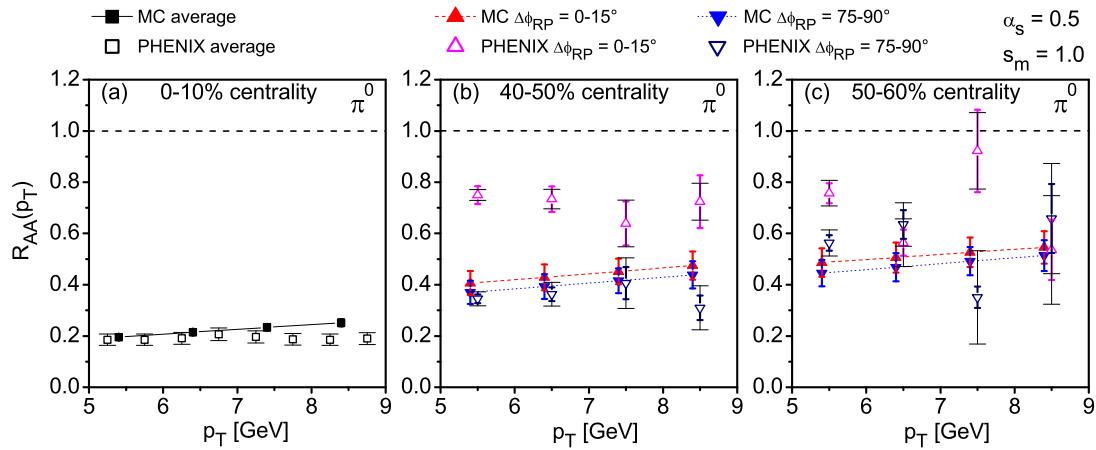}
\vspace{-1.5cm}
\caption{(Color online. Originally published in \cite{Auvinen:2010yt}.) Left panel: The \(\pi^0\) nuclear modification factor for 0-10\% centrality, averaged over the reaction plane angle. Middle and right panel: The \(\pi^0\) nuclear modification factor dependence on the reaction plane angle $\Delta \phi$ for 40-50\% (middle panel) and 50-60\% centrality (right panel). The simulation points (solid squares and triangles) are connected with lines to guide the eye. The PHENIX data are from \cite{PHENIX-R_AA} (0-10\% centrality, open squares) and \cite{PHENIX-R_AA-RP} (40-50\% and 50-60\% centrality, open triangles). Colored bars with small cap represent statistical errors; black bars with wide cap are systematic errors.}
\label{fig_raaphi}
\end{figure*}

As seen from Figure \ref{fig_raaphi}, however, our model cannot reproduce the reaction-plane angle dependence seen in the PHENIX experiment. Also, the computed suppression decreases too slowly as one advances to the more peripheral collisions, which means that the inclusive, angle-averaged nuclear modification factor does not match with the experimental data either. Thus the pathlength-dependent observables strongly disfavor any large contributions from these kind of fully incoherent processes to the energy loss of hard partons. Utilizing our event-by-event hydro, we have also studied the $R_{AA}$ for fluctuating initial state geometry, where the weak sensitivity to the angle-dependent observables is also seen. Overall, our findings, combined with the similar results for radiative energy loss, suggest that the initial state density fluctuations do not play a major role in hard parton energy loss \cite{Renk:2011qi}.

\begin{figure}[htb]
\centering
\includegraphics[width=11.5cm]{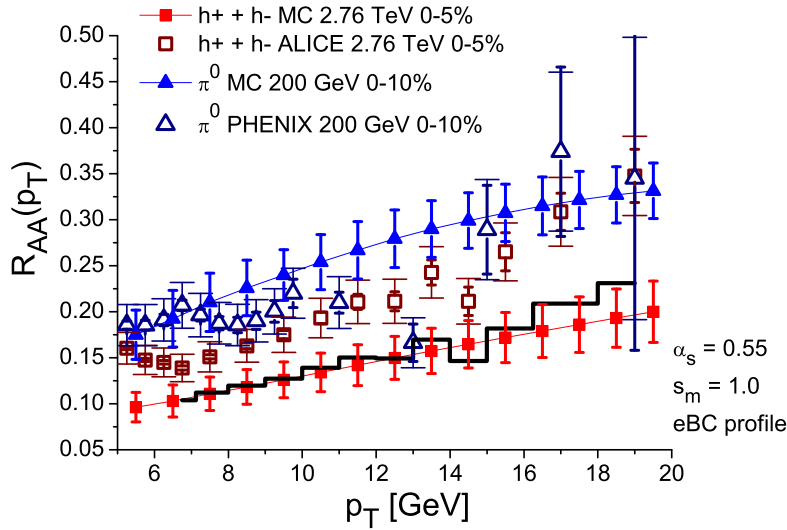}
\vspace{-1.0cm}
\caption{\label{fig_rhiclhc} (Color online) Comparison of simulation and experimental data for neutral pion $R_{AA}(P_T)$ in RHIC $\sqrt{s_{NN}}=200$ GeV Au+Au collisions and charged hadron $R_{AA}(P_T)$ in LHC $\sqrt{s_{NN}}=2.76$ TeV Pb+Pb collisions. The simulation points (solid squares and triangles) are connected with lines to guide the eye. The PHENIX data (open triangles) are from \cite{PHENIX-R_AA} and ALICE data (open squares) are from \cite{Aamodt:2010jd}. Error bars with small cap represent statistical errors; bars with wide cap are systematic errors. The black histogram indicates the lower alternative $pp$ reference for the ALICE data.}
\end{figure}

Figure \ref{fig_rhiclhc} shows how the $R_{AA}(P_T)$ changes as we move from $\sqrt{s_{NN}}=200$ GeV Au+Au collisions at RHIC to $\sqrt{s_{NN}}=2.76$ TeV Pb+Pb collisions at the LHC. Using a (2+1)-dimensional hydro with a smooth eBC profile (as opposed to sWN profile) requires us to re-tune $\alpha_s$ to 0.55 in our model to get the right amount of suppression for neutral pion $R_{AA}(P_T)$ in RHIC energies. Given the large systematic uncertainties of the present data, the $P_T$ dependence of our elastic energy loss model is in reasonable agreement with both the RHIC and published LHC measurements. However, comparison with the preliminary data \cite{Appelshäuser} suggests that the obtained $P_T$ slope at the LHC may be too modest. Fixing the pathlength dependence without changing the $P_T$ dependence too dramatically will be an interesting challenge for future studies.

\ack 
This work was financially supported by grants from the Jenny and Antti Wihuri Foundation (J.A.) and from the Magnus Ehrnrooth Foundation, by the national Graduate School of Particle and Nuclear Physics, and by the Projects 130472 and 13305 of the Academy of Finland. CSC -- IT Center for Science Ltd. is acknowledged for the allocation of computational resources.

\end{document}